\documentclass[10pt,a4paper, twocolumn]{article}
\usepackage[a4paper, left=1in, right=1in, top=1in, bottom=1in, columnsep=20pt]{geometry}
\usepackage[english]{babel}
\usepackage{hyphenat}
\usepackage{microtype}
\usepackage{titlesec}

\titleformat{\section}{\normalfont\normalsize\bfseries} 
  {\thesection}{1em}{}
\titleformat{\subsection}
{\normalfont\small\bfseries}
  {\thesubsection}{1em}{} 

\usepackage{bold-extra}
\usepackage{amsmath}
\usepackage{amssymb}
\usepackage{mathtools} 
\usepackage{braket} 
\DeclarePairedDelimiter\ceil{\lceil}{\rceil}
\usepackage{graphicx}
\usepackage{appendix}
\usepackage{authblk}
\usepackage{csquotes}
\usepackage{xcolor}
\usepackage{etoolbox}
\usepackage{balance}

\usepackage{comment}

\pdfoutput=1
\usepackage{color}
\usepackage{todonotes}
\usepackage{subfigure}
\usepackage{booktabs}
\usepackage{array}
\usepackage{diagbox}
\usepackage{yfonts}
\usepackage{multirow,rotating}
\usepackage{amsmath,amssymb,bm,graphicx,wasysym,mathrsfs,dsfont}
\usepackage{cite}
\usepackage{slashed,relsize}
\usepackage[colorlinks=true,linkcolor=black,anchorcolor=black,citecolor=blue,filecolor=cyan,menucolor=red,runcolor=filecolor,urlcolor=blue,bookmarks=true,bookmarksnumbered=true]{hyperref}
\usepackage{url}
\usepackage{xcolor}
\newcommand{\rev}[1]{\textcolor{black}{#1}}

\usepackage{hyperref}

\usepackage[font=small,labelfont=bf]{caption}

\newcommand{\canc}[1]{}

\def\beq{\begin{equation}\displaystyle\displaystyle}
	\def\eeq{\end{equation}}
\def\bea{\begin{eqnarray}\displaystyle} 
	\def\eea{\end{eqnarray}}

\def\({\left(}
\def\){\right)}
\def\bry{\begin{array}}
	\def\ery{\end{array}}

\makeatletter
\DeclareFontFamily{OMX}{MnSymbolE}{}
\DeclareSymbolFont{MnLargeSymbols}{OMX}{MnSymbolE}{m}{n}
\SetSymbolFont{MnLargeSymbols}{bold}{OMX}{MnSymbolE}{b}{n}
\DeclareFontShape{OMX}{MnSymbolE}{m}{n}{
    <-6>  MnSymbolE5
   <6-7>  MnSymbolE6
   <7-8>  MnSymbolE7
   <8-9>  MnSymbolE8
   <9-10> MnSymbolE9
  <10-12> MnSymbolE10
  <12->   MnSymbolE12
}{}
\DeclareFontShape{OMX}{MnSymbolE}{b}{n}{
    <-6>  MnSymbolE-Bold5
   <6-7>  MnSymbolE-Bold6
   <7-8>  MnSymbolE-Bold7
   <8-9>  MnSymbolE-Bold8
   <9-10> MnSymbolE-Bold9
  <10-12> MnSymbolE-Bold10
  <12->   MnSymbolE-Bold12
}{}

\let\llangle\@undefined
\let\rrangle\@undefined
\DeclareMathDelimiter{\llangle}{\mathopen}%
                     {MnLargeSymbols}{'164}{MnLargeSymbols}{'164}
\DeclareMathDelimiter{\rrangle}{\mathclose}%
                     {MnLargeSymbols}{'171}{MnLargeSymbols}{'171}
\makeatletter
\usepackage{geometry}
\title{Can Quantum Extreme Learning Machines Be \\ Classically Simulated?}
\title{Entanglement and Classical Simulability \\ in Quantum Extreme Learning Machines}

\author{A. De Lorenzis$^{1,2,*}$, \quad M. P. Casado$^{1,3}$, \\
\setlength{\affilsep}{0.2em}
 N. Lo Gullo $^{4,5}$, \quad T. Lux$^1$, \quad F. Plastina$^{4,5}$ and A. Riera$^2$ \\
\setlength{\affilsep}{3em}
\scriptsize
$^{1}$\textit{Institut de Física d’Altes Energies (IFAE) - The Barcelona Institute of Science and Technology (BIST), \\Campus UAB, 08193 Bellaterra (Barcelona), Spain }
\\
$^{2}$\textit{Qilimanjaro Quantum Tech S.L., Carrer de Veneçuela, 74, 08019 Barcelona, Spain} \\
$^{3}$\textit{Departament de Física, Universitat Autònoma de Barcelona }\\
$^{4}$\textit{Dipartimento di Fisica, Università della Calabria, 87036 Arcavacata di Rende (CS), Italy }\\
$^{5}$\textit{INFN, gruppo collegato di Cosenza, 87036 Arcavacata di Rende (CS), Italy}\\
* E-mail: adelorenzis@ifae.es
}
\date{}
\begin{document}
\twocolumn[
    \begin{@twocolumnfalse}
    \maketitle
    \centering
    \vspace{-1cm}
    \noindent\line(2,0){420}\\
    \begin{abstract}
        Quantum Machine Learning (QML) has emerged as a promising framework 
        \rev{for exploring how quantum dynamics may enhance data processing tasks.} Here we investigate Quantum Extreme Learning Machines (QELMs), a quantum analogue of classical Extreme Learning Machines in which training is restricted to the output layer. Our architecture combines dimensionality reduction (via PCA or Autoencoders), quantum state encoding, evolution under an XX Hamiltonian, and projective measurement to produce features for a classical single-layer classifier.\\
        By analyzing the classification accuracy as a function of evolution time, we  \rev{observe} 
        a sharp transition between low- and high-accuracy regimes, followed by saturation. Remarkably, 
        \rev{the saturated performance is comparable to that obtained using Haar-random unitaries} that generate maximally complex dynamics, even though the XX model is integrable and local. \rev{ Our results indicate that this increase in performance } 
        correlates with the onset of entanglement, which improves the embedding of classical data in Hilbert space and leads to more separable clusters in measurement probability space. Thus, 
        \rev{moderate entanglement can contribute positively to the structure of the data representation,} improving learnability without necessarily implying \rev{quantum} computational advantage.\\
        \rev{For the image-classification tasks studied here, namely MNIST, Fashion-MNIST, and CIFAR-10, the relevant evolution time is consistent with information exchange over short distances and, within the explored system sizes, does not show evidence of scaling with the full system size. This suggests that QELM performance in this regime relies only on limited entanglement and remains compatible with efficient classical simulation. Our results clarify how local quantum dynamics and moderate quantum correlations are already sufficient to generate useful feature representations for learning.}
    \end{abstract}
    \noindent\line(2,0){420}\\
    \vspace{1cm}
    \end{@twocolumnfalse}
]
\thispagestyle{empty}
\setcounter{equation}{0}
\setcounter{footnote}{0}
\setcounter{page}{1}
\section{Introduction}\label{sec:intro}
Over the past two decades, quantum computing has evolved from a theoretical concept into a promising technological frontier, with potential applications that span fields such as chemistry, cryptography, finance, and artificial intelligence \cite{feynman1982simulating, Montanaro_2016, Nielsen_Chuang_2010, Shordoi:10.1137/S0097539795293172}. Recent advances in hardware and algorithm development \cite{Arute48651, Zhong:2020iql, Huang:2021pei, Preskill:2018jim} have significantly accelerated progress, fostering renewed interest in practical quantum-enhanced solutions.\\
\rev{A particularly active area of research is Quantum Machine Learning (QML) \cite{Biamonte:2016ugo, book, Schuld:2018gao, Congarticle, Liu:2022bpb}, which explores how quantum-mechanical effects can be used to enrich data representations and learning protocols. In some settings, QML has also been proposed as a possible route to address tasks that may be computationally challenging for classical methods \cite{Schuld2014, Havlicek:2018nqz, Lloyd:2013lby, mastroianni2023assessing, mastroianni2024variational, consiglio2024variational, Vetrano:2024vbh}. More broadly, however, QML provides a useful framework to investigate which features of quantum dynamics may be genuinely relevant for data processing tasks.\\}
\rev{Within this broader context,} we focus on the Quantum Extreme Learning Machine (QELM) \cite{article, Innocenti2023, Xiong, Sakurai:2022ala, Hayashi:2025gwk, DeLorenzis:2024koa, QELMSuprano, Kornjaca:2024afu, chenPhysRevApplied.14.024065, Vetrano:2024vbh, Solanki:2025xji, Hayashi_qer, Sakurai_simpleH}, a quantum adaptation of the classical Extreme Learning Machine (ELM) \cite{1380068,HUANG201532, 10.1007/s10462-013-9405-z, wang2022review, articleHuang2011}. The ELM paradigm simplifies learning by confining optimization to the output layer, and this core idea is preserved in QELMs through the use of a static quantum reservoir, namely a quantum system with fixed internal dynamics. Input data are encoded into quantum states, evolved through the reservoir, and then projected into a high-dimensional feature space, where only the final output weights are trained.\\
\rev{In this work, our goal is not to introduce a new QELM architecture nor to optimize image-classification performance on standard benchmarks. Rather, we use the QELM framework as a controlled setting to investigate which physical features of the underlying quantum dynamics are actually responsible for the emergence of useful feature maps. In particular, we ask whether high performance requires strongly complex or effectively random dynamics, or whether it can already be achieved within a simple local and integrable many-body setting.}\\
\rev{QELMs are computationally appealing because the quantum reservoir itself is not trained iteratively: only the final readout layer is optimized, which substantially reduces the number of trainable parameters compared with conventional deep learning models.\\}
\rev{Moreover, QELMs provide a flexible framework for constructing nonlinear feature maps in high-dimensional Hilbert spaces, which has motivated their use in classification and regression tasks \cite{sakurai22, Hayashi:2022xab}.\\}
\rev{QELMs fall within the broader framework of Quantum Reservoir Computing (QRC) \cite{Fujii2017, AngelatosPhysRevX.11.041062, chenPhysRevApplied.14.024065, Mujal_2021}, while distinguishing themselves by their memoryless nature, where the reservoir dynamics reduce to a static high-dimensional input transformation. More broadly, QRC has been explored across a variety of physical platforms, including spin chains \cite{Xia2023, Sannia2024, Martínez2021, Gotting2023, Martínez2023, Mujal2022, chen2019learning, nakajima2019boosting, tran2020higher, martinez2020information, Settino:2024qvi}, fermionic and bosonic systems \cite{Ghosh2019, HoanTran2023, Llodra2023}, and Rydberg atoms \cite{Kornjaca:2024afu, Bravo2022, Henriet2020, Perciavalle:2025efc, Perciavalle2024}.\\}
\rev{Throughout this paper, we investigate how the performance of a QELM depends on the duration of the quantum evolution when the reservoir is implemented by a local XX Hamiltonian. The choice of this model is motivated by the fact that it provides a minimal and well-controlled many-body setting in which information spreading, entanglement growth, and the conditions for classical simulation can be analyzed transparently.\\
Across different datasets and system sizes within the explored regime, we observe a common qualitative behavior:
\begin{enumerate}
    \item the performance exhibits a relatively sharp transition as a function of the evolution time;
    \item beyond this transition, the accuracy saturates and no significant improvement is obtained for longer evolutions.
\end{enumerate}
These observations suggest that useful feature generation does not require fully complex quantum dynamics, but can already emerge from short-time local evolution and limited entanglement. The rest of the paper is organized as follows: Sec.~2 describes the modelling framework; Secs.~3 and 4 analyze the performance as a function of the evolution time and the onset of the accuracy plateau; Secs.~5 and 6 discuss the transition time and its implications; finally, Sec.~7 presents our conclusions.}\\
\section{Structure and Design of the Adopted Quantum Extreme Learning Machines}\label{sec:The model}
In this work, we investigate a Quantum Extreme Learning Machine 
architecture that integrates a classical Extreme Learning Machine (where only the output layer is subject to training) with a quantum reservoir. A schematic overview \cite{DeLorenzis:2024koa} of the proposed model is provided in Fig.\ref{fig:QELM_model}. The key steps of this approach are:
\begin{figure*}[htbp]
    \centering
\includegraphics[width=0.9\textwidth]{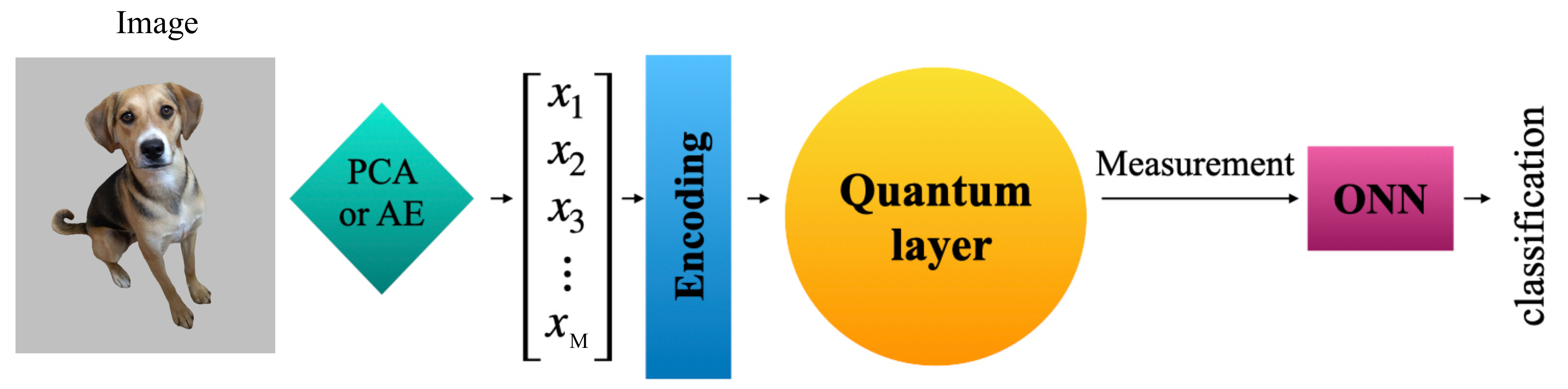}
    \caption{Schematic representation of the QELM architecture. The workflow consists of the following steps: dimensionality reduction using either PCA or an Autoencoder (AE); encoding of the classical data into an initial quantum state; time evolution through the quantum layer; measurement of the evolved quantum state; and final classification via a classical single-layer neural network. Image for illustrative purposes only.}
    \label{fig:QELM_model}
\end{figure*}

\begin{enumerate}
  \item \textit{Feature reduction}: One of the main challenges in applying quantum machine learning lies in the limited number of qubits currently available in quantum hardware. Since many real-world datasets exhibit high-dimensional feature spaces, it becomes essential to employ dimensionality reduction techniques that retain as much of the original information as possible while adapting the data to the constraints of the quantum reservoir. To this end, we compress the input data into a compact latent representation before feeding it into the quantum model. This reduction is achieved through two distinct approaches: Principal Component Analysis (PCA) and Autoencoders (AEs) \cite{KRAMER1992313, Hinton1993AutoencodersMD, Rumelhart1986LearningRB,doi:10.1126/science.1127647}.\\
  PCA is a linear statistical method that projects the original features onto a set of orthogonal components, ranked by their variance, thereby preserving the most informative directions in the data. In contrast, Autoencoders are nonlinear neural networks trained to reproduce the input at the output by passing it through a bottleneck structure. The encoding process maps the data into a lower-dimensional latent space, which is then reconstructed by the decoder.
  \item \textit{Quantum Encoding of Classical Data}: Before quantum processing can take place, classical data must be encoded into quantum states that can be handled by the quantum system. In our approach, this is done by initializing each qubit in a pure state derived from the input features.\\
  A single qubit's pure state can be visualized as a point on the Bloch sphere:
  \begin{equation} \label{eqn:blochsphere}
    \ket{\psi} = \cos (\frac{\theta}{2})\ket{0} + e^{ i \phi} \sin (\frac{\theta}{2})\ket{1}
\end{equation}
  where $\theta \in [0,\pi]$ and $\phi \in [0,2\pi]$. This geometric representation enables the encoding of real-valued data by mapping feature values to these two angles.\\
  \rev{In our case, the original image data are first processed by the classical dimensionality-reduction stage (PCA or Autoencoder, depending on the dataset), which produces a latent feature vector of dimension $M=2N$.}\\
  We adopt a dense angle encoding strategy, where each qubit is initialized using two classical features: one mapped to the polar angle $\theta$ and the other to the azimuthal angle $\phi$.\\
  The complete initial quantum state of the system, composed of $N = \ceil*{M / 2}$ qubits for a \rev{latent} feature vector $\vec{x} = [x_1, ..., x_M]^T \in \mathbb{R}^M$ is given by:
{\small
\begin{equation} \label{eqn:denseangleencoding}
    \ket{\vec{x}} = \bigotimes_{i=1}^{\ceil*{M / 2}} \left( \cos (\frac{x_{2i -1}}{2})\ket{0} + e^{ i x_{2i}} \sin(\frac{x_{2i -1}}{2})\ket{1}\right) 
\end{equation}
}\\
\rev{Before encoding, the components of $\vec{x}$ are normalized feature-wise using the minimum and maximum values computed on the training set, and then rescaled to the interval $[0,\pi]$. The first $N$ components are assigned to the polar angles $\theta_i$, while the remaining $N$ components are assigned to the azimuthal angles $\phi_i$. \\
We adopt this encoding because it provides a simple and hardware-efficient embedding of classical data into the quantum state space while keeping the encoding stage fixed and free of additional trainable parameters. This choice is also supported by our previous comparative study of QELM encodings and preprocessing strategies \cite{DeLorenzis:2024koa}, in which dense angle encoding showed consistently good performance across the considered datasets. This is important for the purpose of the present work, since it allows us to isolate the role of the subsequent quantum dynamics. At the same time, we note that assigning different geometric roles to the two components entering each qubit state may introduce representation-dependent biases, which are not the focus of the present work.\\
Here, each pair of features ($x_{2i -1}$, $x_{2i}$) is used to set the state of the i-th qubit, effectively allowing each qubit to encode two features. This encoding efficiently maps high-dimensional classical inputs into a quantum state suitable for further processing, while respecting the qubit limitations of current hardware.} 
  \item \textit{Quantum layer and time evolution}: Once the classical data has been encoded into a quantum state, the system undergoes time evolution governed by a Hamiltonian operator. This evolution, carried out within the quantum layer, transforms the initial state through unitary dynamics. In our implementation, the system evolves under the 
  \rev{XX Hamiltonian with nearest-neighbor coupling:}
  \begin{equation}\label{eq:XXmodel}
 H= \frac{1}{2} \sum_{i=1}^{N} (\sigma_x^{(i)}\sigma_x^{(i+1)} +\sigma_y^{(i)}\sigma_y^{(i+1)}).
\end{equation}
\rev{The XX Hamiltonian is chosen here not as a generic black-box reservoir, but as a minimal and well-controlled many-body model with local interactions. This makes it possible to relate the learning performance to physically meaningful dynamical processes, such as short-range information spreading and entanglement generation, and to assess whether such ingredients are already sufficient to produce useful feature maps. Moreover, because the XX model is integrable, it provides a controlled setting in which the relation between evolution time, correlation growth, and classical simulability can be investigated transparently.}
  \item \textit{Measurement and Classification}: After the quantum evolution governed by the selected Hamiltonian, the final state is measured in the computational basis. \rev{The relevance of this basis choice in relation to the Hamiltonian is discussed in Appendix~B.} This process yields classical information corresponding to the probability distribution over basis states, i.e., the populations of the wave function.\\
  In a real quantum experiment, such information is typically obtained by repeatedly measuring identically prepared quantum states, due to the probabilistic nature of quantum measurements. 
  \rev{In the results presented in this work, however, the measurement probabilities are computed directly from the full statevector in the computational basis, corresponding to the ideal infinite-sampling limit. Therefore, the reported performances characterize the feature map generated by the quantum dynamics itself, without additional fluctuations due to finite-shot measurement statistics. In a realistic implementation, finite-shot sampling would introduce statistical noise in the estimated measurement probabilities, which may affect the quality of the resulting classical features and, consequently, the classification performance.}
  \rev{The resulting classical feature vector contains $2^N$ real values for each input, where $N$ is the number of qubits.} 
  These features are then fed into a classical simple One-layer Neural Network (ONN) with softmax activation, which performs the final classification. \rev{As in Extreme Learning Machines, the quantum reservoir is kept fixed and is not trained iteratively; in our implementation, only the final classical readout layer is optimized, using the Adam optimizer with categorical cross-entropy loss.}  The architecture of the classifier is defined by the number of quantum measurement outputs and the number of target classes. 
  For the numerical implementation, we used the Keras framework.
\end{enumerate}
\subsection{Datasets}
\begin{figure*}[htbp]
    \centering
\includegraphics[width=0.9\textwidth]{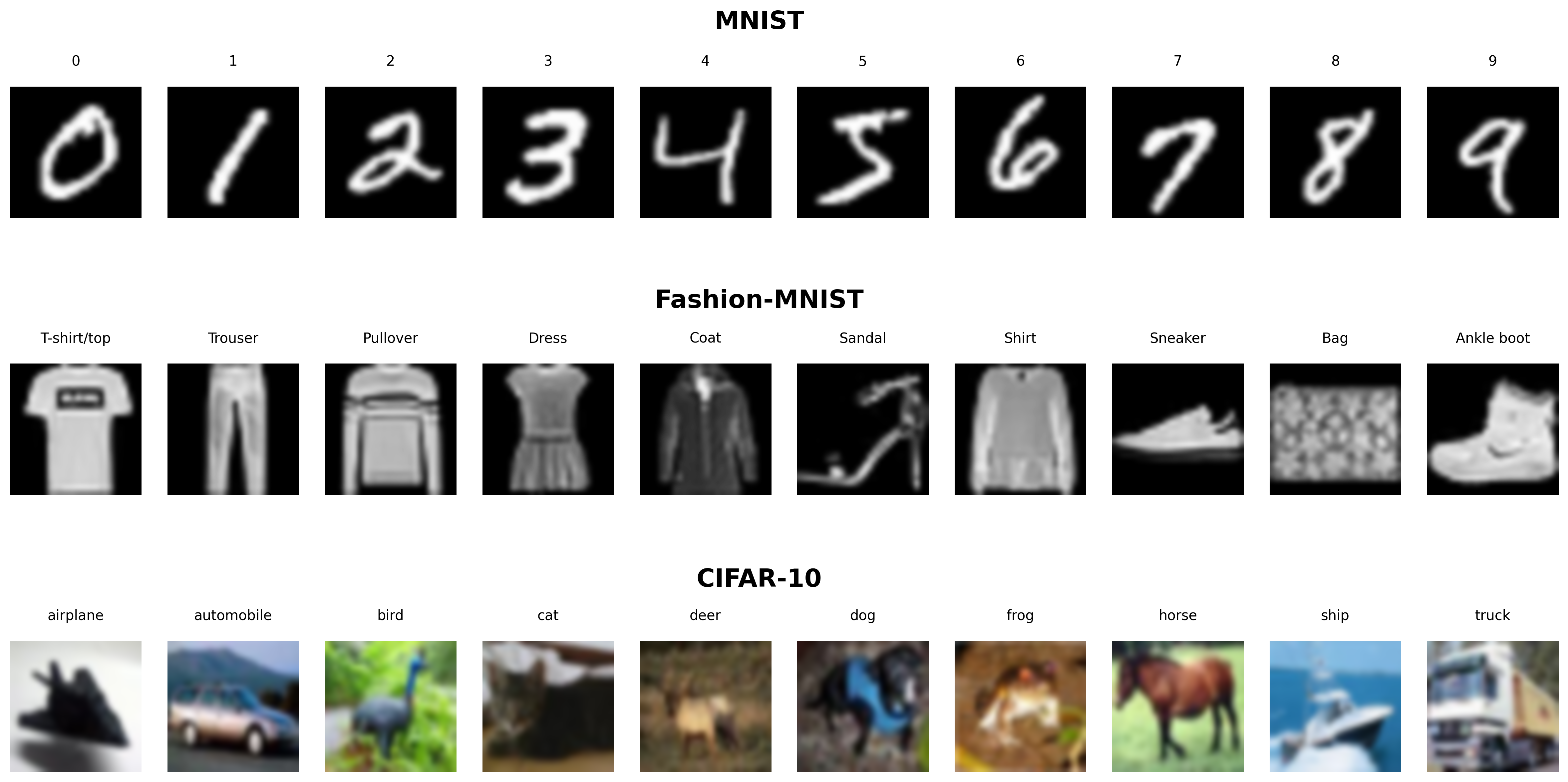}
    \caption{Sample images from three widely used benchmark datasets in this study. The top row displays handwritten digits from MNIST (digits 0–9), the middle row shows clothing items from Fashion-MNIST (10 clothing categories), and the bottom row presents natural object categories from CIFAR-10. Each column corresponds to one class, labeled above each image.}\label{fig:Datasets_sample}
\end{figure*}

\rev{The datasets considered in this work are not chosen to emphasize the application itself or to compete with state-of-the-art classical methods, but rather to provide controlled and widely used benchmarks on which the dependence of QELM performance on the underlying quantum dynamics can be assessed systematically. Accordingly, in this section we describe the benchmark datasets used to train and evaluate our QELM model: MNIST \cite{mnist}, Fashion-MNIST \cite{fashionmnist} and CIFAR-10 \cite{cifar10}.}\\
Figure~\ref{fig:Datasets_sample} provides a visual overview of these datasets, with one representative image per class.\\
\begin{itemize}
  \item MNIST: The MNIST (Modified National Institute of Standards and Technology) dataset is a classical benchmark for handwritten digit classification. It contains 70,000 grayscale images of 28×28 pixels, split into 60,000 training and 10,000 test samples. Each image represents a digit from 0 to 9.
  \item Fashion-MNIST: Fashion-MNIST is a drop-in replacement for MNIST, designed to provide a more challenging classification task. It consists of 70,000 grayscale images of clothing items, also sized 28×28 pixels, divided into 10 categories such as shirts, shoes, and coats. Like MNIST, it provides 60,000 training and 10,000 test images.
  \item CIFAR-10: CIFAR-10 is a dataset of 60,000 color images (RGB) at 32×32 pixel resolution, distributed equally among 10 classes, including objects such as airplanes, cats, and trucks. It is split into 50,000 training and 10,000 test images. Due to the diversity and visual complexity of the images, CIFAR-10 serves as a standard benchmark for evaluating image classification algorithms in more realistic settings.
\end{itemize}
\section{Performance as a Function of Evolution Time}
\begin{figure*}[htbp]
    \centering
    \subfigure[MNIST, training]{
        \includegraphics[width=0.45\textwidth]{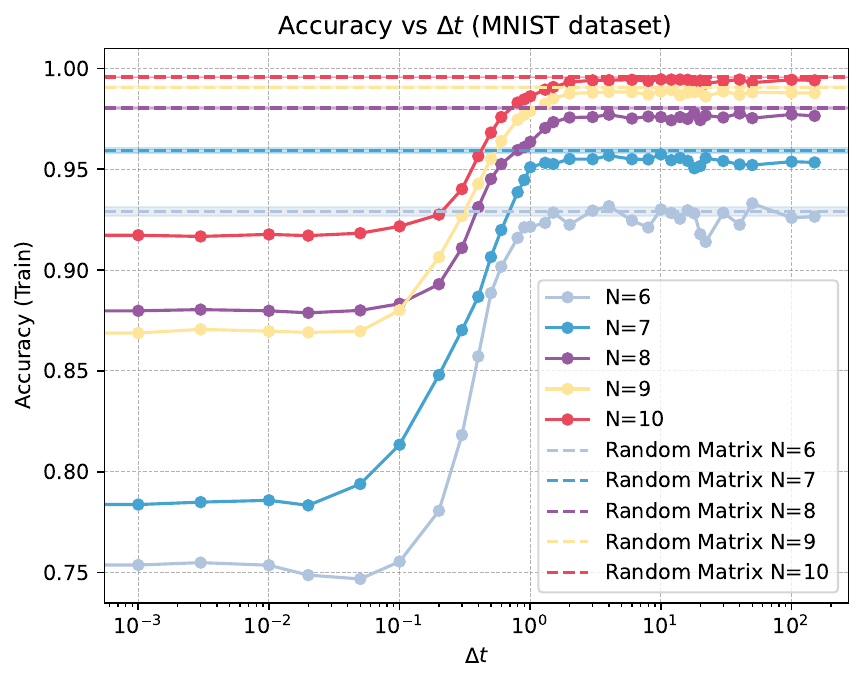}
    }
    \hfill
    \subfigure[MNIST, test]{
        \includegraphics[width=0.45\textwidth]{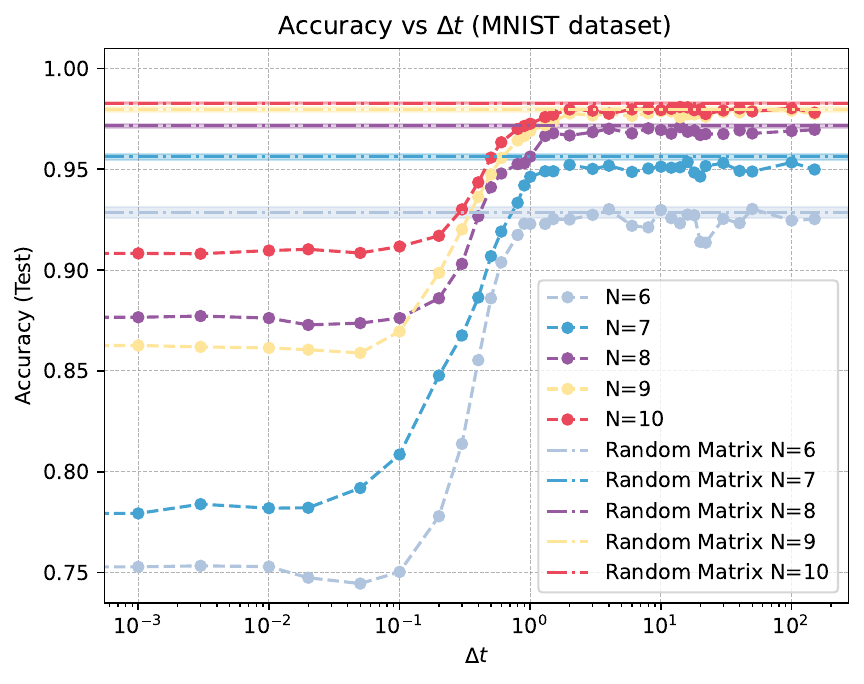}
    }

    \vspace{0.3cm}

    \subfigure[Fashion-MNIST, training]{
        \includegraphics[width=0.45\textwidth]{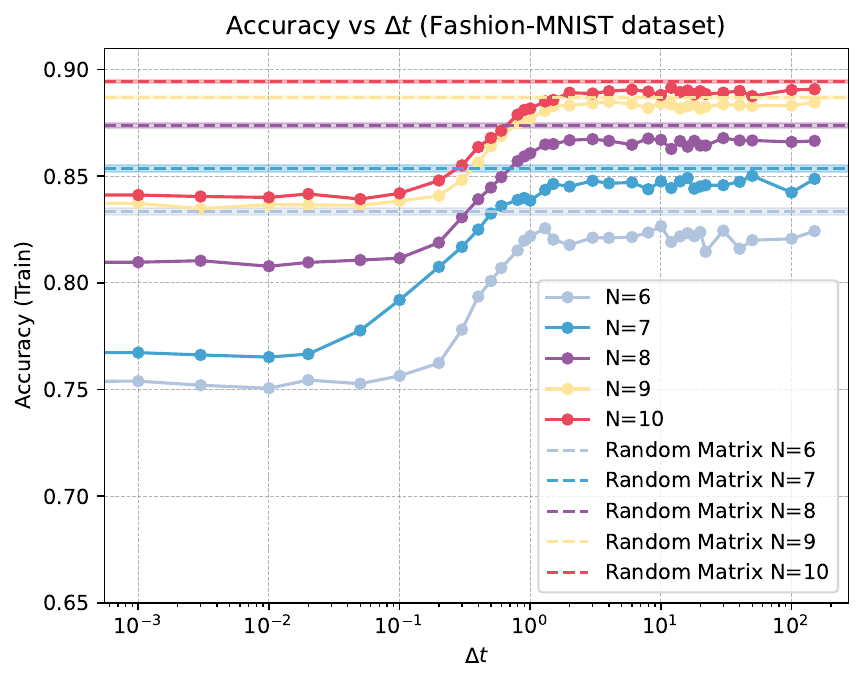}
    }
    \hfill
    \subfigure[Fashion-MNIST, test]{
        \includegraphics[width=0.45\textwidth]{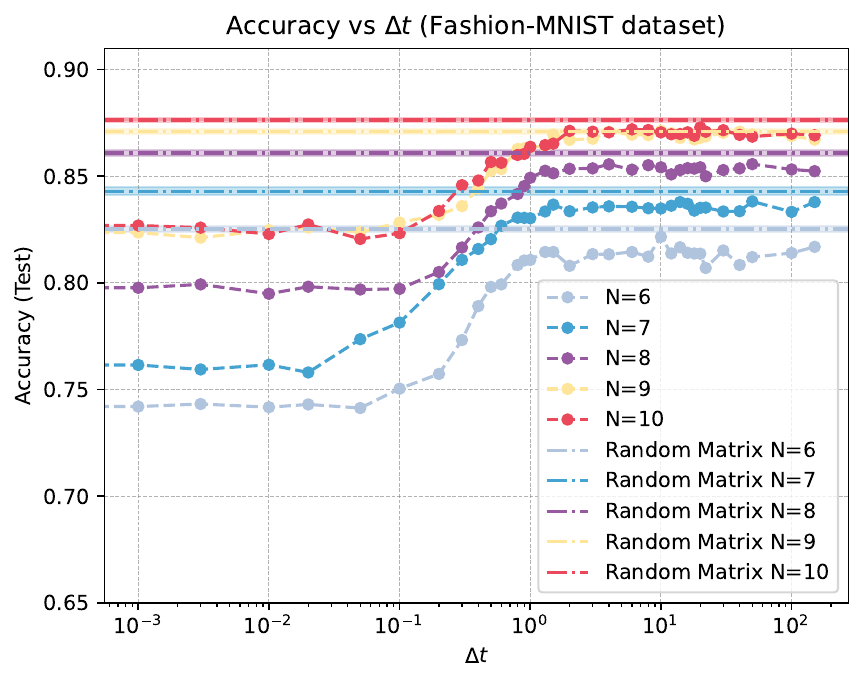}
    }

    \vspace{0.3cm}

    \subfigure[CIFAR-10, training]{
        \includegraphics[width=0.45\textwidth]{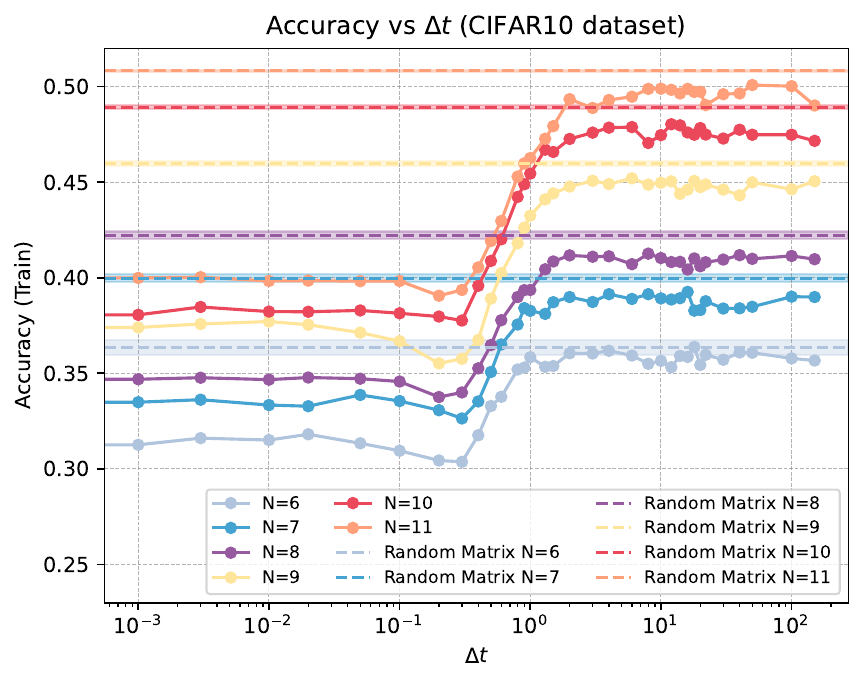}
    }
    \hfill
    \subfigure[CIFAR-10, test]{
        \includegraphics[width=0.45\textwidth]{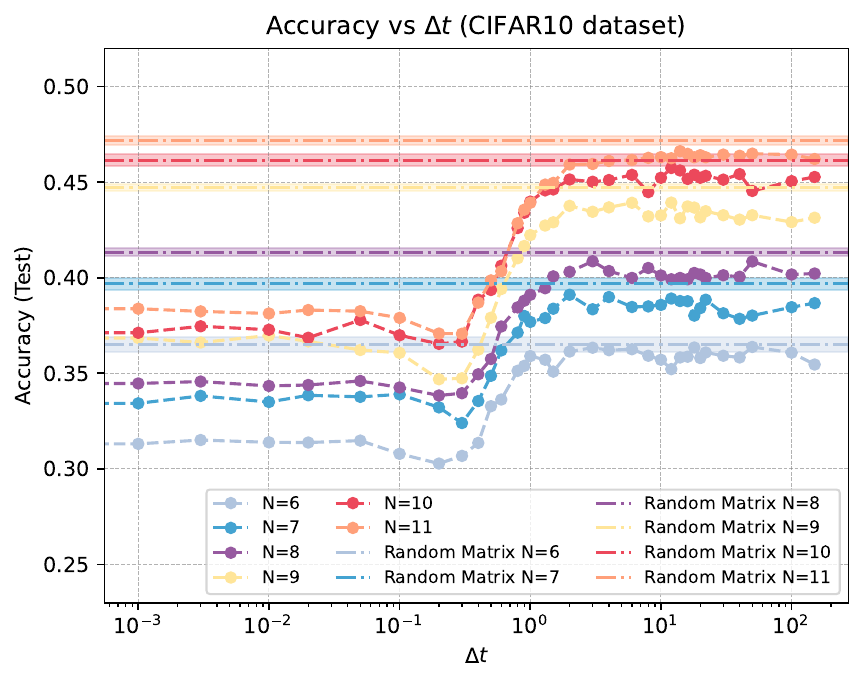}
    }

    \caption{Training (left column) and testing (right column) accuracy as a function of evolution time for the MNIST (top row), Fashion-MNIST (middle row), and CIFAR-10 (bottom row) datasets. The time evolution and the encoding have been performed with the XX Hamiltonian and dense angle encoding, respectively. The horizontal dashed lines indicate the performance obtained when the XX evolution operator is replaced by a Haar-random unitary acting globally on the full $N$-qubit Hilbert space. For each system size, results are averaged over 10 independent realizations, and the shaded area represents the corresponding standard deviation.}
    \label{fig:Accuracy_all_datasets}
\end{figure*}

\rev{We stress that the purpose of these benchmarks is not to compete with highly optimized classical machine-learning architectures, which can achieve substantially higher accuracies on the same datasets, but rather to investigate how the performance of the QELM depends on the underlying quantum dynamics. Accordingly, the main focus of the present analysis is not the absolute benchmark accuracy, but the relation between evolution time, feature generation, and classification performance.}\\
After reducing the dataset images to $2N$ features and encoding them using dense angle encoding, we evolved the system and performed the measurements. These outcomes were then fed into a single-layer neural network, which provided us with the resulting classification accuracies.\\ We performed the classification and calculated the accuracies for different evolution times and studied the performance in terms of achieved accuracy as a function of evolution time. \\ In Fig. \ref{fig:Accuracy_all_datasets} we can see these plots respectively for the MNIST, Fashion-MNIST and CIFAR-10 datasets, with the training results shown on the left and the test results on the right. The colors of the curves represent the different number of qubits employed in the algorithm. For the simpler datasets, such as MNIST and Fashion-MNIST, we employed between 6 and 10 qubits, whereas for the more complex dataset CIFAR-10, results were plotted for 6 to 11 qubits. Simulating quantum computers with many qubits requires substantial computational resources, so our study was limited to systems of manageable size.\\
The figures show that each curve remains nearly constant at first. After a certain time interval, the accuracy increases sharply and eventually converges to a plateau. There is thus a relatively narrow transitional time window during which the performance rises until it stabilizes. 
\rev{Within the explored system sizes and datasets, we observe that the performance reaches a steady regime after a relatively short evolution time $t^*$, without clear evidence that this timescale grows with the full system size.}
\\
The curves are shifted upward as N increases, which is expected since the number of features used in the algorithm grows as $2N$.\\
In sum, we observe that there is a transition time $t_*$ at which the accuracy exhibits a sudden jump, and a saturation accuracy $A_*$ that the system reaches after the transition time $t_*$.\\
In the following sections, we aim to explain the reason behind these values of saturation accuracy $A_*$ and transition time $t_*$.

\section{Information Propagation and the Accuracy Plateau}
Recently, Ref.\cite{Vetrano:2024vbh} proposed a connection between the performance of QELM and the phenomenon of information scrambling. In quantum information, \rev{scrambling can be understood as the delocalization of initially local information into highly nonlocal correlations, which may render it inaccessible to simple local observables.}
\rev{If the information becomes strongly delocalized, its relation to the input may become inaccessible to simple observables or low-complexity measurements, even though it is not fundamentally lost in the unitary dynamics.}
However, the impact of scrambling depends strongly on the observables under consideration. 
\rev{When focusing on local quantities, such as magnetization or correlation functions, scrambling is related to the thermalization of subsystems, which can progressively reduce the accessibility of the correlations between the encoded input data and the measured outcomes.}
As the size of the subsystem increases, the effect of scrambling diminishes. In the limit where the subsystem approaches the size of the entire system, it becomes possible to recover a mapping between the input and the output by measuring global observables.
In our framework, these global observables correspond to the probabilities of finding the system in specific bit-string states (computational basis states). The recovery of the input–output mapping, despite local scrambling, arises because the dynamics of global observables remain unitary. 
\rev{Thus, although a Hamiltonian that induces system-wide information spreading may reduce the amount of information accessible through local observables, it can actually enhance input distinguishability when probed through global measurements.}
\rev{To test this idea, we consider Haar-random unitary matrices acting on the full Hilbert space, which serve as reference models of maximally mixing quantum dynamics. Such random unitaries provide a natural benchmark, as they maximize the spreading of information from the initial state across the entire system.}\\
\rev{In Fig.~\ref{fig:Accuracy_all_datasets} (horizontal dashed lines), we show the performance of the QELM when the XX evolution operator is replaced by a Haar-random unitary acting globally on the full $N$-qubit state. Since different Haar-random unitaries can lead to different accuracies, the dashed lines correspond to the average over 10 independent realizations sampled from the Haar measure on $U(2^N)$. The associated standard deviation is indicated by the shaded area. The different system sizes are represented by different colours, consistent with the colour code used for the Hamiltonian evolution.}\\
\rev{This comparison is physically relevant because it allows us to assess whether useful feature generation requires fully complex, effectively structureless quantum dynamics, or whether it can already be achieved within a simple local and integrable model.}
\rev{We find that the saturation performance $A_*$ obtained for each explored system size is comparable to the performance achieved by a Haar-random unitary, in agreement with the intuition that classification benefits from information spreading.}\\
This observation is particularly remarkable, given that the XX Hamiltonian employed in the QELM is:
\begin{itemize}
    \item Highly specific and the evolution operator it generates is far from random,
    \item Translational invariant,
    \item Integrable, as it can be diagonalized through a Jordan--Wigner transformation followed by a Fourier transform.
\end{itemize}
\canc{Despite being both constrained and integrable, from the perspective of accuracy in classification tasks, the XX Hamiltonian effectively emulates the action of a random unitary, thereby realizing dynamics that are indistinguishable from maximally scrambling evolution.}
The explanation lies in the structure of the encoding scheme. In this setup, the initial state significantly overlaps with nearly all many-body eigenstates of the XX model. As a result, the dynamics is not confined to integrable sectors but  effectively explores the entire Hilbert space.
\begin{figure*}[htbp]
    \centering
    \subfigure[]{
        \includegraphics[width=0.45\textwidth]{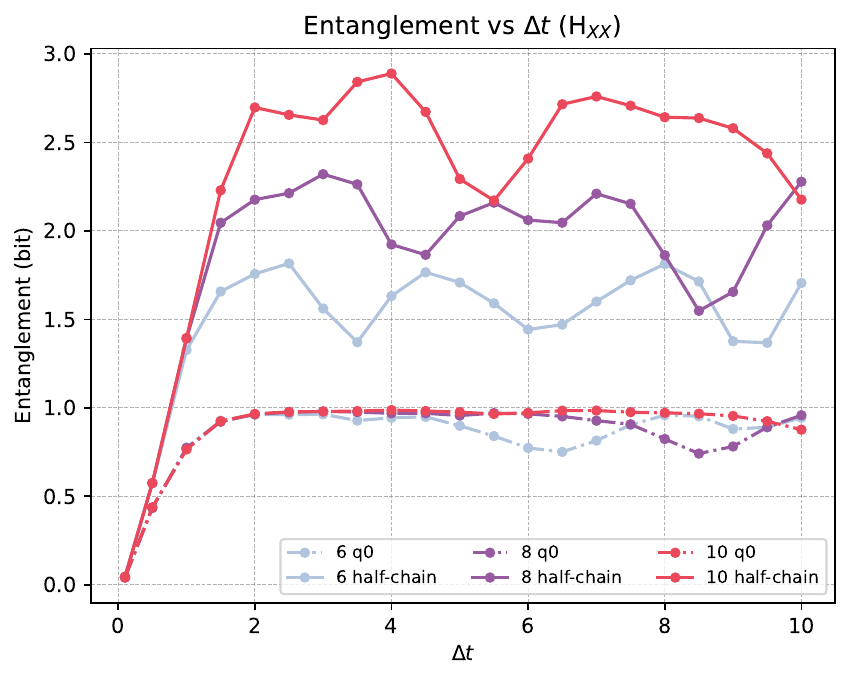}
    }
     \hspace{0.001\textwidth}
    \subfigure[]{
        \includegraphics[width=0.45\textwidth]{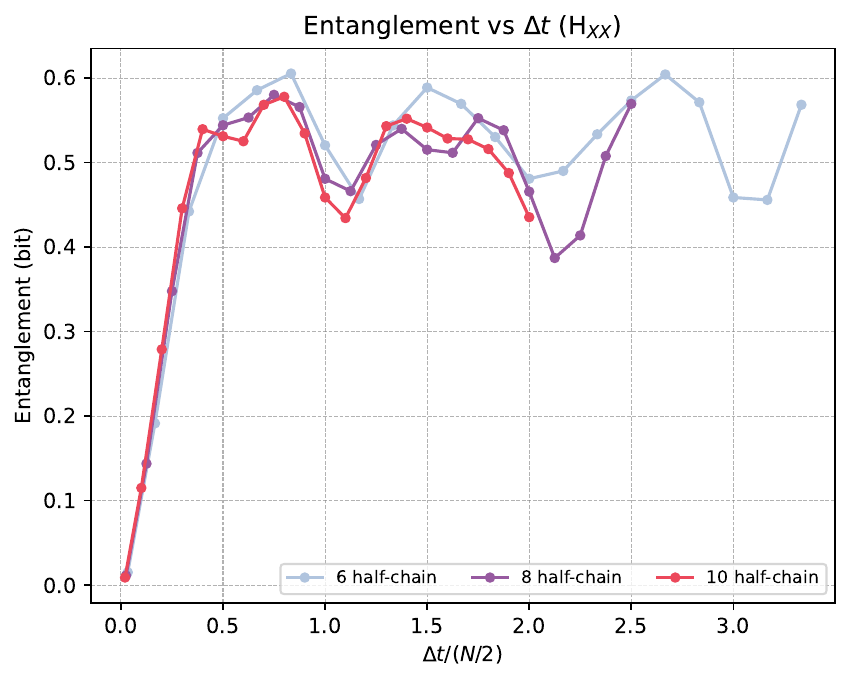}
    }
    \caption{a) Von Neumann entropy as function of the evolution time of half-chain (solid lines) and of the single qubit (dash-dotted lines). The colors denote the different sizes $N$ ($N=6, 8, 10$) of the entire system. b) Normalized (with $N/2$) von Neumann entropy  as function of the evolution time (divided by $N/2$) of half-chain.}
    \label{fig:Entanglement}
\end{figure*}

\section{Critical Transition Time and Locality}

Once we have understood the saturation value $A_*$ of the accuracy, let us now address the question of the transition time $t_*$.\\
A natural first guess, consistent with the intuition of information spreading across the system, is that the transition time corresponds to the characteristic timescale required for initially localized information on individual qubits to spread throughout the system.\\
To substantiate this intuition, it is useful to introduce the notion of an effective speed of information propagation. Although in non-relativistic quantum mechanics space and time are treated on different footings, and super-luminal influences of local perturbations are, in principle, possible, the Lieb--Robinson bounds demonstrate that for local Hamiltonians---such as our XX model---there exists an effective causal structure. In practice, this establishes a maximum velocity at which quantum dynamics can propagate information.\\
In Appendix A we determine the Lieb--Robinson velocity for the XX model defined in Eq.~\eqref{eq:XXmodel} and show that it is equal to $1$. Combining this result with the plots presented in the previous section, two conclusions follow: 
\begin{itemize}
    \item \rev{From a qualitative perspective, the data do not provide clear evidence that the transition time grows linearly with the system size within the explored range, suggesting that the onset of useful feature generation may not require information to spread across the entire system.}
    \item Quantitatively, up to the transition time $t_*$, the information remains confined to the qubit of origin and its nearest neighbors.
\end{itemize}
To better visualize how information spreads across the system and to confirm the above insights, in Fig.~\ref{fig:Entanglement}(a) we plot the von Neumann entropy of half the chain together with the single-qubit von Neumann entropy as functions of the evolution time. These quantities provide complementary perspectives: the half-chain entropy captures the build-up of bipartite entanglement across the system, while the single-qubit entropy measures the delocalization of local information.\\
The evolution of the single-qubit entropy shows that each qubit, where information initially encoded, becomes entangled with its environment after an evolution time of order $1$, irrespective of the total system size. In contrast, the entanglement entropy of half the chain grows linearly with time until it saturates. The growth rate is identical for all system sizes, indicating that the speed of information propagation is independent of system size, while the saturation value scales proportionally to the block size ($N/2$). These results on the entanglement dynamics of spin chains are well established in the literature~\cite{Calabrese_2005,Latorre_2009} and are explained by the ballistic propagation of quasiparticle excitations, which travel in opposite directions in entangled pairs.\\
It is also interesting to note that the curves of half-chain entropy presented in Fig.~\ref{fig:Entanglement}(a) obey a scaling law of the form  
\begin{equation}
    S_N(t) = N\, f\!\left(\tfrac{t}{N}\right)\, ,
\end{equation}
where $S_N(t)$ denotes the half-chain entropy for a system of $N$ sites at time $t$. One could expect this relation to hold for large $N$; but, in fact, it approximately holds even for not so large numbers of sites. Indeed, the scaling function $f$ is shown in Fig.~\ref{fig:Entanglement}(b), where the normalized entropy $S_N/(N/2)$ is plotted as a function of $t/N$ for the system sizes $N=6, 8$, and $10$.\\
Several features of $f$ are worth highlighting:
\begin{itemize}
    \item The entropy grows linearly until it saturates at $t = N/2$. 
    \item For $t > N/2$, the entropy exhibits oscillations with characteristic periodicities $T \simeq 0.5$ and $T \simeq 1$.
\end{itemize}
These properties of the scaling function $f$ confirm the intuition of ballistic information propagation. The linear growth for $t < N/2$ arises from pairs of entangled quasiparticles located on opposite sides of the bipartition: the longer the time, the more quasiparticles cross its boundary and contribute to the entropy \cite{Calabrese_2005}. Once the block is filled with quasiparticles, the entropy saturates, since quasiparticles are simultaneously entering and leaving. The oscillations observed for $t > N/2$ are a consequence of all quasiparticles propagating with similar velocities but carrying different amounts of entanglement.\\
The confirmation of the ballistic information propagation model, together with the observation of a transition time $t_*$ at which the information has only spread across a local region of a few neighbors, suggests that our naive real-space picture of information spreading is not the key feature for explaining the information processing capacity of the QELM. This indicates that the analysis should instead be carried out in a different space.
\section{The Probability Polytope}
\begin{figure*}[htbp]
    \centering
\includegraphics[width=0.5\textwidth]{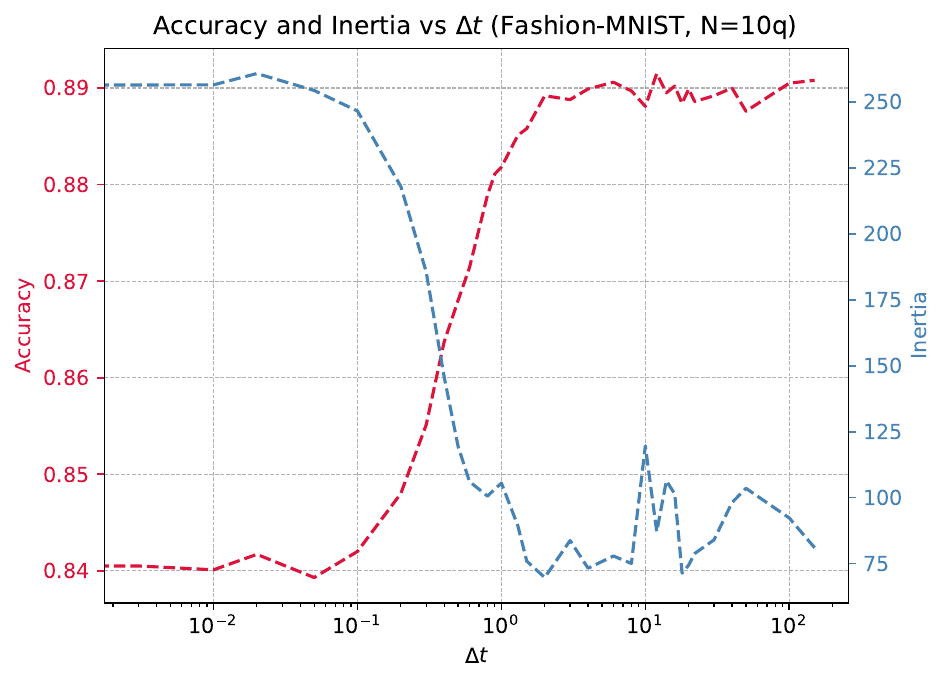}
    \caption{Accuracy (in magenta color) and Inertia (in light blue color) as a function of evolution time using the Fashion-MNIST dataset and $N=10$.}
    \label{fig:Inertia}
\end{figure*}

\begin{figure*}[htbp]
    \centering
    \subfigure[]{
        \includegraphics[width=0.45\textwidth]{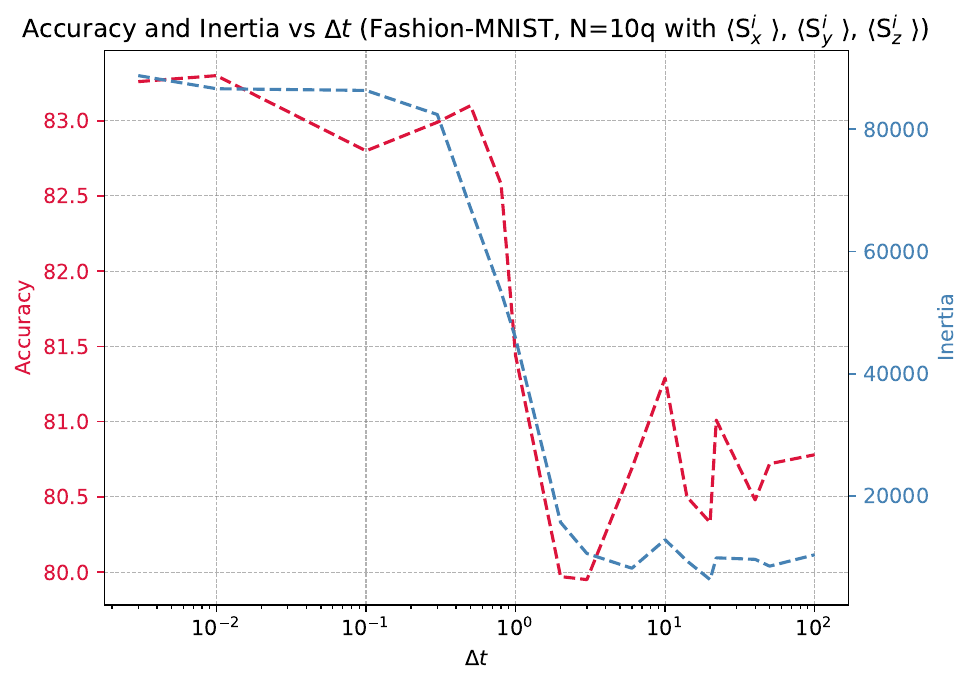}
    }
     \hspace{0.001\textwidth}
    \subfigure[]{
        \includegraphics[width=0.45\textwidth]{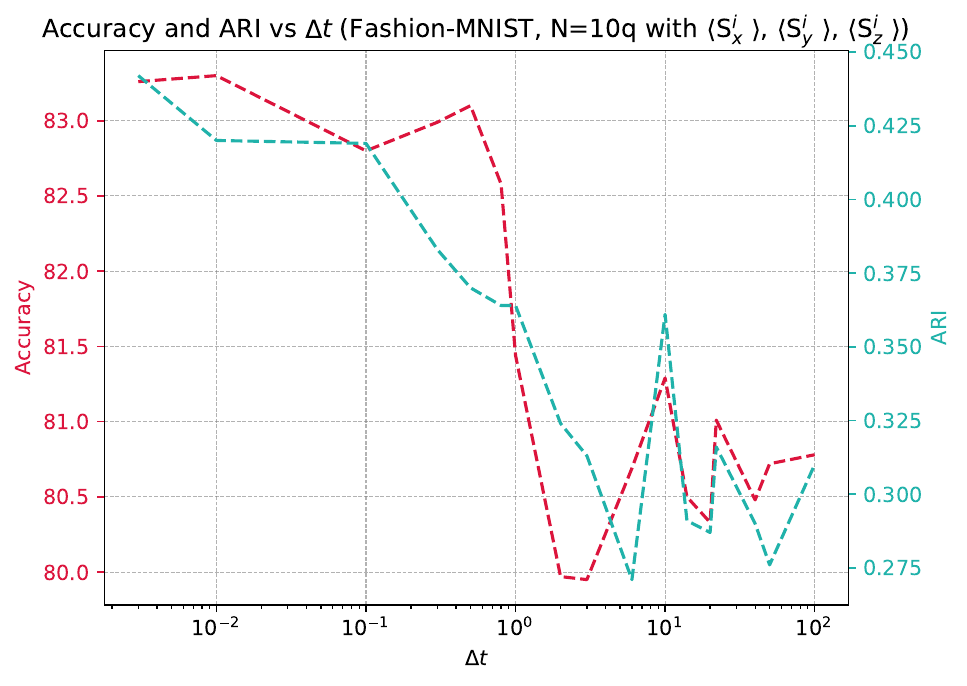}
    }
    \caption{a) Accuracy (in magenta color) and Inertia (in light blue color) as a function of evolution time using the Fashion-MNIST dataset and $N=10$.  b) Accuracy (in magenta color) and ARI (in light green color) as a function of evolution time using the Fashion-MNIST dataset, for $N=10$, obtained from the local expectation values only. \rev{Here, ``local'' refers to the set of single-qubit Pauli expectation values $\{\langle \sigma_i^x \rangle, \langle \sigma_i^y \rangle, \langle \sigma_i^z \rangle\}_{i=1}^N$.}}
    \label{fig:Expectation_values}
\end{figure*}

A natural framework for analyzing the dynamics of the Quantum Extreme Learning Machine is the \emph{probability polytope}, defined by the set of all possible probability distributions over the computational basis states. More precisely, we consider vectors of real numbers $\vec{p} \in \mathbb{R}^{2^N}$ whose $2^N$ components satisfy
\[
p(s) \geq 0 \quad \forall s \in \{0,1\}^N, 
\qquad \sum_{s \in \{0,1\}^N} p(s) = 1,
\]
where each component is indexed by a bit string of length $N$, $s = (s_1, s_2, \ldots, s_N)$.  
The probability vector $\vec{p}$ associated with a quantum state $\ket{\psi}$ is given by
\begin{equation}
  p(s)=\left|\bra{s}\psi\rangle\right|^2,
\end{equation}
where $\ket{s}$ denotes the computational basis element corresponding to the bit string $s$.\\
The motivation for introducing the probability space is that it directly corresponds to the feature space on which the classical neural network operates. Consequently, it provides the most natural setting in which to study how the quantum evolution reshapes the data representation accessible to the classical classifier.\\
Because the probability polytope has exponentially many dimensions, analyzing the temporal evolution of data points within it is far from trivial.  
To gain qualitative insight into the structure and distribution of the emerging features, we applied the K-means clustering algorithm. This unsupervised learning method reveals natural groupings among the data points, thereby exposing latent structure and relationships within the feature space.\\
The K-means algorithm minimizes the within-cluster sum of squared distances between samples and their corresponding centroids. It proceeds iteratively by assigning each data point to the nearest centroid (assignment step) and updating the centroid positions as the mean of their assigned members (update step). The process continues until convergence, defined as the point at which cluster assignments stabilize or the change in the objective function falls below a predefined threshold.\\
Figure~\ref{fig:Inertia} shows the evolution of the \emph{inertia}—defined as the within-cluster sum of squared distances—which quantifies cluster compactness.  
As the evolution time increases, the inertia exhibits a consistent decrease, indicating that the feature vectors become more tightly clustered and better separated.  
This trend parallels the increase in classification accuracy observed in the output neural network (ONN): as the inertia decreases, the ONN accuracy rises, revealing an inverse correlation between these two quantities.\\
To substantiate our previous discussion about local information scrambling, we also computed both accuracy and inertia using local information only. 
\rev{Namely, we consider the single-qubit Pauli expectation values
$\langle \sigma_i^x \rangle$, $\langle \sigma_i^y \rangle$, and $\langle \sigma_i^z \rangle$
for $i=1,\dots,N$, obtaining the plot shown in Fig. \ref{fig:Expectation_values} (a).}
We observe that, 
\rev{unlike in} the previous analysis based on the string probabilities, the ONN accuracy and the inertia \rev{now} follow a similar trend and \rev{both} decrease over time.  Notice, in particular, that the decrease occurs on a time of the order of $t^*$. \rev{Interestingly, the slight decrease in the local accuracy and inertia appears to start already before $t^*$. This behavior is consistent with a progressive delocalization of the information initially encoded in local observables under the XX dynamics: as the evolution proceeds, local measurements become less informative, while the full probability distribution gradually develops the global structure required for successful classification. Around $t^*$, this structure becomes sufficiently pronounced to yield well-separated clusters in the full measurement space. This accuracy drop indicates that local information becomes progressively less accessible on this timescale, consistent with a delocalization of the encoded information under the dynamics.} 
\rev{Furthermore, the plot shows that, in this case, the inertia is not a reliable tracker of the algorithm's performance: although the data do cluster in this locally extracted latent space, the resulting clusters are no longer aligned with the underlying class structure. To gain deeper insight, we further evaluated the Adjusted Rand Index (ARI), which measures the similarity between the clustering predicted by the K-means algorithm and the actual class labels. As shown in Fig.~\ref{fig:Expectation_values}(b), the ARI decreases over time consistently with the accuracy. This behavior confirms that the ability of the ONN to predict the correct class deteriorates as the clustering induced by local observables becomes progressively less informative.}\\
\rev{As a summary of this discussion: while the data cluster correctly in the full $\mathbb{R}^{2^N}$ space, with the accuracy increasing and the inertia decreasing after $t^*$, when considering only the $3N$-dimensional space of the local Pauli expectation values, the clustering is no longer aligned with the class structure, showing that local observables alone are no longer sufficient to extract the relevant information.}
\section{Implications for Classical Simulability}
\begin{figure*}[htbp]
    \centering
\includegraphics[width=0.5\textwidth]{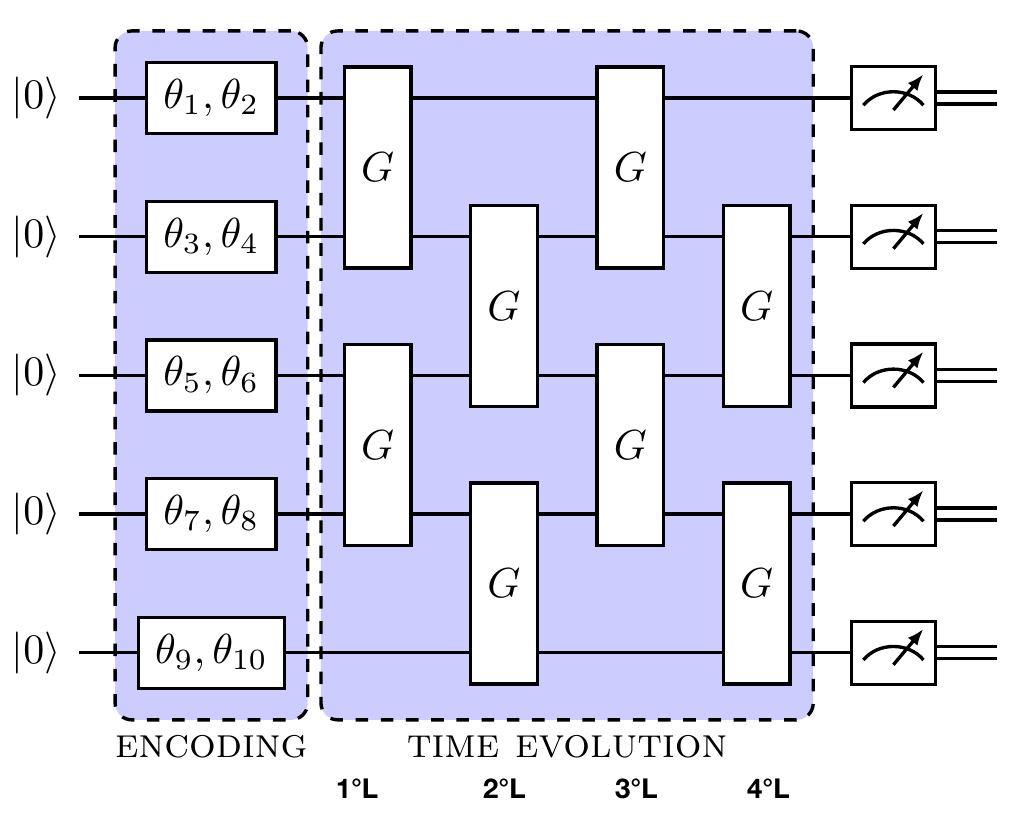}
\caption{A schematic representation of an algorithm with 5 qubits, in which a unitary random matrix is applied for pairs of qubits contiguous.}\label{fig:Schematic}
\end{figure*}

\begin{figure*}[htbp]
    \centering
    \subfigure[MNIST, training]{
        \includegraphics[width=0.45\textwidth]{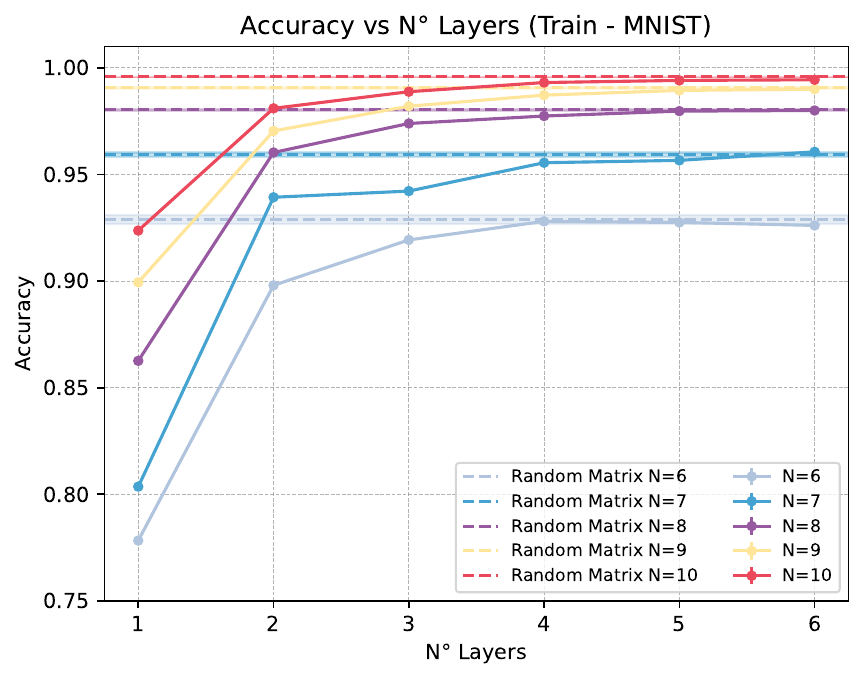}
    }
    \hspace{0.01\textwidth}
    \subfigure[MNIST, test]{
        \includegraphics[width=0.45\textwidth]{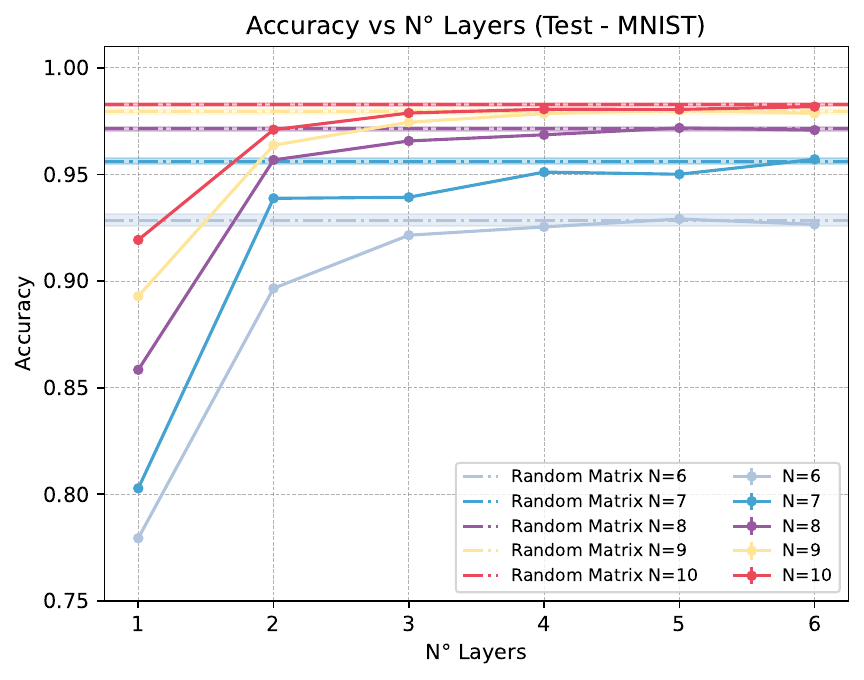}
    }

    \vspace{0.2cm}

    \subfigure[CIFAR-10, training]{
        \includegraphics[width=0.45\textwidth]{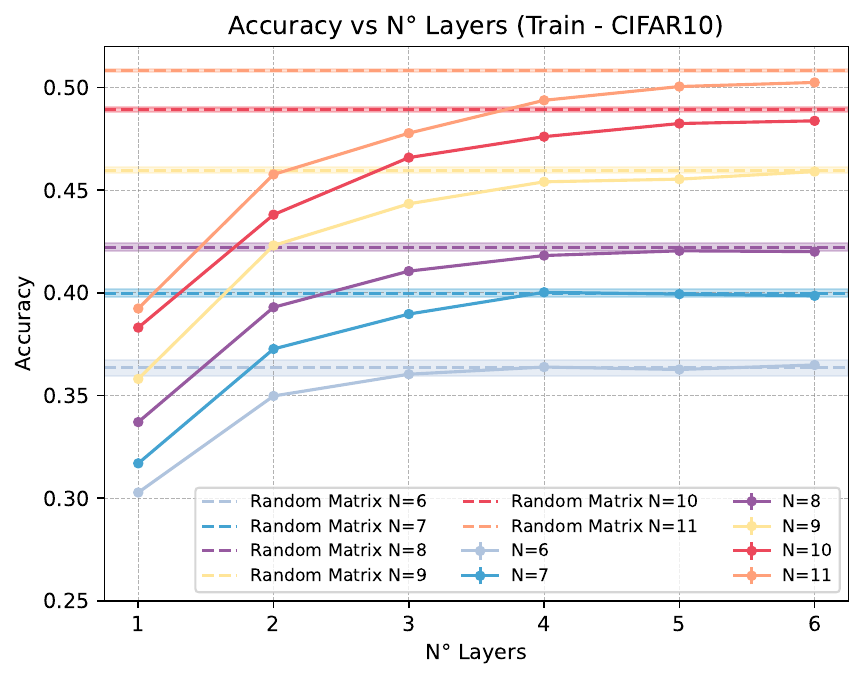}
    }
    \hspace{0.01\textwidth}
    \subfigure[CIFAR-10, test]{
        \includegraphics[width=0.45\textwidth]{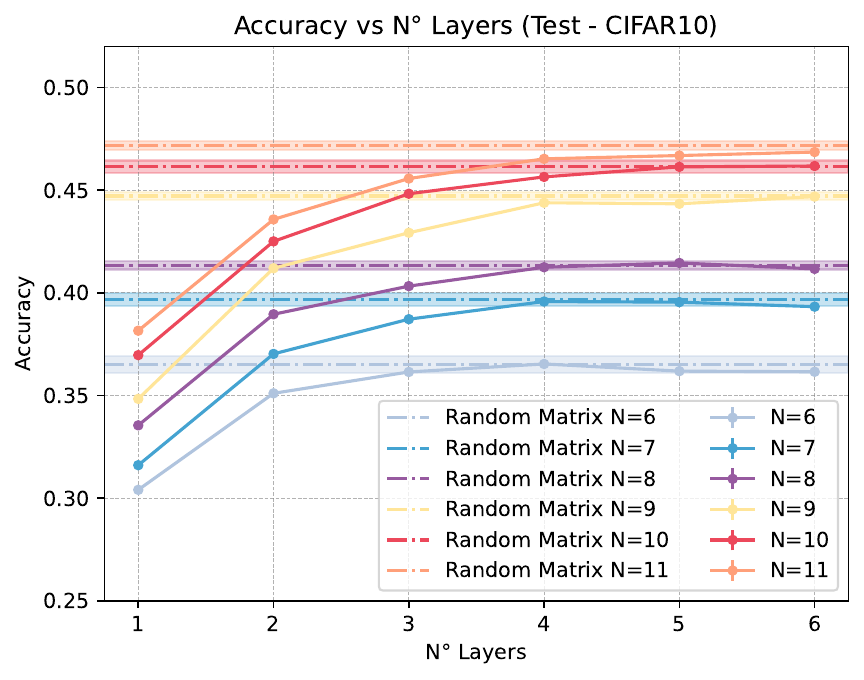}
    }

    \caption{Training (left column) and testing (right column) accuracy as a function of the number of layers for the MNIST (top row) and CIFAR-10 (bottom row) datasets.}
    \label{fig:Accuracy_layers}
\end{figure*}

\rev{Our results indicate that optimal accuracy does not require information from a single site to spread across the entire system.}
Instead, the propagation of information to the nearest neighbors already suffices to extract the relevant features needed for learning.\\
\rev{The observation that the maximum accuracy is reached on a timescale that, within the explored system sizes, does not show evidence of scaling with the full system size has important implications for the classical simulability of QELMs.}
In particular, it indicates that the effective depth of the quantum evolution is shallow, and that the relevant correlations remain short-ranged. 
\rev{Such shallow circuits, characterized by limited entanglement growth, are compatible with efficient classical simulation using tensor-network techniques such as Matrix Product States (MPS). From this perspective, the learning process in the regime studied here does not necessarily demand computational resources beyond classical capabilities.}\\
To further substantiate these ideas, we implemented an alternative, explicitly digital version of the model (see Fig.~\ref{fig:Schematic}). Using the same reduced features employed as input for the QELM and maintaining the same dense-angle encoding of the initial states, we applied a sequence of two-qubit random unitaries $G$ acting on contiguous qubit pairs. At each successive layer, the unitary is shifted by one qubit, as illustrated schematically in Fig.~\ref{fig:Schematic}.\\
This construction ensures that, at each layer, information carried by each qubit is mixed only with that of its immediate neighbor. As the depth increases, correlations propagate across larger regions of the system following a well-defined light-cone structure.\\
We applied this algorithm to the MNIST and CIFAR-10 datasets and evaluated the classification accuracy as a function of the number of layers (see Fig.~\ref{fig:Accuracy_layers}). In all cases, maximum accuracy was achieved without requiring information to spread throughout the entire system. For the MNIST dataset, four layers were sufficient to reach the accuracy plateau \rev{within the explored
range of system sizes.} A similar trend was observed for CIFAR-10 at smaller $N$, while for larger systems the plateau was reached after the fifth or sixth layer.
\rev{This mild dependence on system size suggests that both the depth required to reach saturation and the corresponding degree of classical simulability may be problem-dependent.}\\


\vspace{-10pt}

\section{Conclusions and Outlook}\label{sec:conclusion}
\rev{In this work, we used Quantum Extreme Learning Machines (QELMs) as a controlled setting to investigate which physical features of the underlying quantum dynamics are responsible for the emergence of useful feature maps for classification.}
We first reduced the dataset images (via PCA or AE) into vectors of $2N$ features, encoded them using the Dense Angle encoding in the quantum layer, and then performed the temporal evolution of the quantum state under the Hamiltonian $H_{XX}$. After the evolution, we measured the output state in the computational basis, obtaining $2^N$ real values for each image, which were then fed into a classical single-layer neural network classifier.\\
Plotting the performance as a function of the evolution time, we observed a relatively sudden transition from a low-accuracy to a high-accuracy regime, followed by saturation. 
\rev{The saturated performance is comparable to that obtained using Haar-random unitaries, which provide a reference model of maximally complex dynamics.}
Remarkably, this equivalence is achieved at relatively short evolution times, when the XX Hamiltonian has generated only local entanglement between neighboring qubits. Despite its simplicity and  integrability, \rev{the XX model is thus able to reach a classification performance comparable to that of a fully random unitary evolution,}
which induces maximal scrambling across the Hilbert space. This finding underscores that complex global dynamics are not necessary to obtain rich and expressive feature maps: even structured, physically realistic interactions can provide sufficient diversity in the quantum embeddings for effective learning.\\
Our analysis further shows that the \textit{improvement in accuracy coincides with the onset of entanglement}, indicating that quantum correlations enhance the embedding of classical data into the Hilbert space. As entanglement ``kicks in,'' the encoded samples become more separable and better clustered in probability space, leading to improved classification performance. \rev{In this sense, the onset of entanglement correlates with a more effective data embedding, providing a physically meaningful interpretation of the transition dynamics observed in QELMs.}\\
\rev{Interestingly, the critical transition time is consistent with information propagating only over short distances and, within the explored system sizes, does not show evidence of scaling with the full system size.}
This observation implies that QELMs rely primarily on local quantum correlations to achieve high accuracy. \rev{Consequently, while quantum resources clearly enhance feature extraction, the limited depth and moderate entanglement observed in this regime suggest compatibility with efficient classical simulation, though larger-scale studies would be needed to assess the generality of this conclusion.}\\
Beyond these observations, our analysis of data evolution within the probability polytope provides a geometric interpretation of the accuracy transition. As the quantum dynamics unfold, the measurement probability vectors exhibit increasingly structured and class-distinguishable patterns, which enhance the discriminative power of the classical classifier. Thus, the gain in performance arises not from global information spreading, but from the \textit{emergence of organized and class-separable manifolds in Hilbert space}—a process through which moderate quantum correlations bridge quantum dynamics and classical feature learning.\\
\rev{Overall, our results show that useful feature generation in QELMs can already emerge from local, integrable, and moderately entangling dynamics, without requiring fully complex quantum evolution. This makes QELMs a valuable setting in which to study the interplay between quantum dynamics, learnability, and classical simulability, and to probe more sharply the boundary between classically accessible regimes and genuine quantum advantage.}
\section*{Acknowledgments}
We would like to thank Marcin Plodzien for helpful discussions.
The research leading to these results has received funding from the Spanish Ministry of Science and Innovation PID2022-136297NB-I00 /AEI/10.13039/501100011033/ FEDER, UE and through the Spanish State Research Agency, under Severo Ochoa Centres of Excellence Programme 2025-2029 (CEX2024001442-S) funded by MICIU/AEI/10.13039/501100011033. IFAE is partially funded by the CERCA program of the Generalitat de Catalunya.
The results have received funds from the program Plan de Doctorados Industriales of the Research and Universities Department of the Catalan Government (2022 DI 011). This work was partially funded by the Generalitat de Catalunya, AGAUR 2021-SGR-01506 and partially supported by the
PNRR MUR Project No. PE0000023-NQSTI through the
secondary projects QuCADD and ThAnQ. We acknowledge the use of the Finnish CSC facilities under the project 2010295 "Quantum Reservoir Computing".
We gratefully acknowledge the use of RES resources provided by the Barcelona Supercomputing Center through project FI-2025-1-0043 on MareNostrum 5.
\begin{appendices}
\section{Appendix: Lieb-Robinson Bounds: General Statement}
Consider a quantum spin system defined on a lattice $\Lambda$, with local Hilbert space $\mathcal{H}_i$ at each site $i \in \Lambda$, and a Hamiltonian of the form:
\begin{equation}
    H=\sum_{X \subset \Lambda} \Phi(X),
\end{equation}
where $\Phi(X)$ are local interaction terms supported on finite subsets $X \subset \Lambda$.\\
We assume that:
\begin{itemize}
    \item The local interaction terms $\Phi(X)$ are uniformly bounded: $\|\Phi(X)\| \leq J$ for some constant $J>0$.
    \item The interactions are of finite range $r_0$ or decay exponentially with distance.
\end{itemize}
Under these assumptions, the Lieb-Robinson bound asserts that for any two observables A and B supported on disjoint regions $X, Y \subset \Lambda$, the following holds:
\begin{equation}
    \|[A(t), B]\| \leq c \|A\|\|B\| e^{- \mu (d(X, Y)-v|t|)},
\end{equation}
where $c$, $\mu$ and $v$ are positive constants. The Lieb-Robinson velocity $v$ depends linearly on the bound $J$ of the interaction terms and scales as $v \propto J$.\\
This captures the idea that stronger local interactions permit faster information propagation across the system.\\

\subsection{Application to the XX Model}
The one-dimensional XX model describes nearest-neighbor spin interactions:
\begin{equation}
\begin{split}
H=J \sum_{j=1}^{N} (\sigma_j^x \sigma_{j+1}^x+\sigma_j^y \sigma_{j+1}^y)=\\
J\sum_{j=1}^{N} (\sigma_j^+ \sigma_{j+1}^-+\sigma_j^- \sigma_{j+1}^+)
\end{split}
\end{equation}
where $\sigma_j^{\pm}=(\sigma_j^x \pm i \sigma^y_j)/2.$\\
Using the Jordan-Wigner transformation , this model maps to free fermions:
\begin{equation}
    H=J\sum_{j=1}^{N} (c_j^\dagger c_{j+1}+ c_{j+1}^\dagger c_j)
\end{equation}.\\
Since the Hamiltonian is quadratic, the Heisenberg evolution $c_j(t)$ remains linear:
\begin{equation}
    c_j(t)=\sum_k u_{jk}(t) c_k, u_{jk}(t)=i^{j-k} J_{j-k}(2Jt),
\end{equation}
where $J_n$ is the Bessel function of the first kind.\\
Therefore, the commutator between distant fermionic operator is:
\begin{equation}
    \|\{c_j(t), c_k^\dagger\}\|=u_{jk}(t)=|J_{j-k}(2Jt)|\leq C e^{- \mu (|j-k|-2Jt)},
\end{equation}
which explicitly realizes the Lieb-Robinson bound with velocity:
\begin{equation}
    v_{LR}=2J.
\end{equation}

\subsection{Group Velocity and its Connection to Lieb-Robinson Bounds}
The free-fermion Hamiltonian can be diagonalized with Fourier modes:
\begin{equation}
    c_j=1/\sqrt{N} \sum_k e^{ikj} a_k,
\end{equation}
with $k=2 \pi m /N$ with $m \in \mathbb{Z}_N$. The Hamiltonian becomes:
\begin{equation}
    H=\sum_k \epsilon(k) a_k^\dagger a_k, \epsilon(k)=2J \cos(k).
\end{equation}
where $\epsilon(k)$ is the dispersion relation.\\
The group velocity is then given by:
\begin{equation}
    v_g(k)=\frac{d \epsilon(k)}{dk}= -2 J sin(k).
\end{equation}
Its maximum over k $\in [-\pi, \pi]$ reads
\begin{equation}
    v_{max}=\max_{k} |v_g(k)| =2J=v_{LR}.
\end{equation}
Thus, the Lieb-Robinson velocity coincides with the maximal group velocity of the system's excitations.

\subsection{Information Propagation from a Localized Excitation}
\begin{figure}
    \centering
    \includegraphics[width=0.5\linewidth]{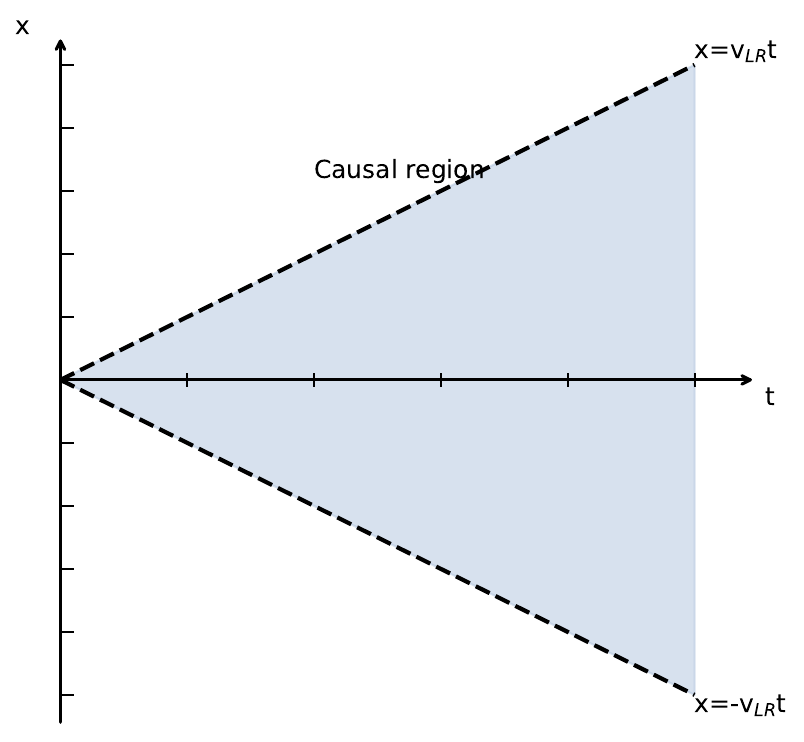}
    \caption{Lieb-Robinson light cone with velocity $v_{LR}=2J$. Outside this region, commutators of local observables are exponentially suppressed.}
    \label{fig:cone}
\end{figure}

Consider a localized excitation created at t=0:
\begin{equation}
    |\psi(0)\rangle = c_j^\dagger |0 \rangle.
\end{equation}
Its time evolution:
\begin{equation}
    |\psi(t)\rangle = c_j^\dagger(t) |0 \rangle
    = \sum_k i^{k-j} J_{k-j}(2Jt)\, c_k^\dagger |0\rangle ,
\end{equation}
shows that the amplitude on site k is $|J_{k-j}(2Jt)|$, which is peacked around $|k-j|\approx 2Jt$, and decays exponentially outside.\\
This illustrates a ballistic propagation of the excitation with a velocity $v=2J$ and confirms that information spreads within a light cone determined by the Lieb-Robinson bound(see Fig.\ref{fig:cone} and \cite{Amico}).
\subsection{Accuracy saturation, Lieb-Robinson velocity and quantum advantage}
In Figures \ref{fig:Accuracy_all_datasets} we show how the accuracy of a QELM depends on the total quantum evolution time for different system sizes and problems. As expected, the accuracy increases with time until it saturates at a value that depends on the size of the quantum reservoir:larger systems achieve higher saturation accuracy.\\
Two complementary observations can be drawn from this plot.\\
First, from a qualitative perspective, \rev{within the explored range of system sizes, the saturation time appears to remain on a relatively short timescale, without clear evidence of growth proportional to the full system size.}\\
Second, from a quantitative standpoint, we can relate this behavior to how far information spreads within the system. 
Notably, all system sizes reach their maximum accuracy at approximately time t=1. According to the Lieb-Robinson bound derived in the previous section, this corresponds to quantum information propagating over a distance of about one lattice site,
\begin{equation}
    d=v_{LR} \cdot t = (2J)t=2 \cdot \frac{1}{2} \cdot 1 = 1.
\end{equation}

\section{Appendix: The choice of Measurement Basis}
In addition to allowing the QELM sufficient time to evolve, it is also crucial to consider the choice of the measurement basis. In particular, the measurement basis should be sufficiently distinct from the eigenbasis of the Hamiltonian governing the evolution. This distinction plays a significant role in the expressiveness and information content of the output features extracted via measurement.\\
\subsection{Quantum Evolution and Basis Representations}
Assume the quantum system underlying the QELM evolves under a time-independent Hamiltonian H, with spectral decomposition:
\begin{equation}
    H=\sum_k E_k |E_k\rangle \langle E_k|,
\end{equation}
where ${|E_k\rangle}$ are the eigenstates and ${E_k}$ the corresponding eigenvalues.\\
The features of each input data sample are encoded into an initial quantum state $|\psi(0)\rangle$, which can be expanded in the Hamiltonian eigenbasis as:
\begin{equation}
    |\psi(0)\rangle= \sum_n c_n |E_n\rangle.
\end{equation}
After an evolution time t, the system evolves into:
\begin{equation}
    |\psi(0)\rangle=\sum_n c_n e^{-iE_n t} |E_n \rangle.
\end{equation}
Let the measurement be performed in a basis ${|s_k\rangle}$, which may or may not coincide with the eigenbasis of H. The probability of measuring the system in state $|s_k\rangle$ at time t is given by:
\begin{equation}
\begin{split}
    p_k(t)=|\langle s_k| \psi(t)\rangle|^2=\\\sum_{n,m} \langle s_k|E_n\rangle \langle s_k|E_m\rangle^* e^{-i(E_n-E_m)t}c_n c_m^*.
\end{split}
\end{equation}
\subsection{Relevance of Basis Misalignment}
This expression illustrates that the time-dependent measurement statistics $p_k(t)$ are sensitive to the overlap between the Hamiltonian eigenstates and the measurement basis vectors. If the two bases coincide, time evolution does not generate any non-trivial interference terms, and the dynamics may lack the richness needed for expressive feature extraction.
The following scenarios clarify this relationship.

\subsection{Identical Hamiltonian and Measurement Bases}
If the measurement basis ${|s_k\rangle}$ coincides with the Hamiltonian eigenbasis ${|E_k\rangle}$, then all cross terms n m in $p_k(t)$ vanish due to the orthogonality. Consequently, the probabilities $p_k(t)$ become time-independent:
\begin{equation}
    p_k(t)=|c_k|^2 \text{(constant in time)}.
\end{equation}
In this case, no interference occurs, and the feature map is static, rendering the QELM ineffective in capturing temporal dynamics.

\subsubsection{Mutually Unbiased Bases}
When the Hamiltonian and measurement bases are mutually unbiased, that is,
\begin{equation}
    |\langle s_k|E_n\rangle|^2=\frac{1}{d} \forall n, k,
\end{equation}
where d is the dimension of the Hilbert space. In this scenario, the probability amplitudes are evenly distributed, and the evolution induces maximal interference among eigenstate components. As a result, the measurement probabilities $p_k(t)$ exhibit rich time-dependent patterns, enhancing the nonlinear feature encoding capabilities of the QELM.\\

\end{appendices}

\bibliographystyle{ieeetr}
\bibliography{bibliography}
\end{document}